\documentclass[useAMS,usenatbib]{mn2e}
\usepackage{graphicx,graphics,amsmath,amssymb,mathrsfs}

\newcommand{\half}{\frac{1}{2}}

\newcommand{\beq}{\begin{equation}}
\newcommand{\eeq}{\end{equation}}

\newcommand{\bfr}{\mbox{\boldmath $r$}}
\newcommand{\bfv}{\mbox{\boldmath $v$}}
\newcommand{\prate}{\mbox{$\dot{\varpi}$}}
\newcommand{\del}{\mbox{\boldmath $\nabla$}}

\newcommand{\p}{\mbox{$\partial$}}
\newcommand{\cals}{{\cal S}}
\newcommand{\cald}{{\cal D}}
\newcommand{\xs}{{x_{\star}}}
\newcommand{\ks}{{k_{\star}}}
\newcommand{\gs}{{\Gamma_{\star}}}

\begin{document}
\title {Slow $m=1$ instabilities of softened gravity Keplerian discs}

\author[Sridhar \& Saini]{S. Sridhar$^{1,3}$ and Tarun Deep Saini$^{2,4}$   \\
  $^{1}$ Raman Research Institute, Sadashivanagar, Bangalore, India, 560 080 \\
  $^{2}$ Indian Institute of Science, Bangalore, India, 560 012 \\
  $^{3}$ ssridhar@rri.res.in 
  $^{4}$ tarun@physics.iisc.ernet.in \\
} 
\maketitle

\begin{abstract}
We present the simplest model that permits a largely analytical
exploration of the $m=1$ counter--rotating instability in a ``hot''
nearly Keplerian disc of collisionless self--gravitating matter. The
model consists of a two--component softened gravity disc, whose linear
modes are analysed using WKB. The modes are {\it slow} in the sense
that their (complex) frequency is smaller than the Keplerian orbital
frequency by a factor which is of order the ratio of the disc mass to
the mass of the central object. Very simple analytical expressions are
derived for the precession frequencies and growth rates of local
modes; it is shown that a nearly Keplerian disc must be
unrealistically hot to avoid an overstability. Global modes are
constructed for the case of zero net rotation.
\end{abstract}


\section{Introduction}

Galactic nuclei are thought to harbour supermassive black holes and
dense clusters of stars, whose structural and kinematical properties
appear to be correlated with global galaxy properties
\citep{geb96, fm00, geb00}. The imprint of galaxy formation is expected to be 
recorded in the nature of stellar orbits. A remarkable case is that of
our nearest large neighbouring galaxy M31, whose centre has a
double--peaked distribution of stars
\citep{lds74, lau93, lau98, kb99}. \citet{tre95} proposed that the
off--centered peak marks the region, in a disc of stars, where lie the
apoapses of many eccentric orbits. Self--gravitating models of such an
eccentric disc have been proposed \citep{bac01, ss01, ss02}. Of
particular interest to the present investigation is the model proposed
in \citet{ss02}, because it included a few percent of stars on
retrograde (i.e. counter--rotating) orbits. Here, it was proposed that
the lopsidedness of the nuclear disc of M31 could have been excited by
the presence of the retrograde stars, which were accreted to the
centre of the galaxy in the form of a globular cluster that spiraled
in due to dynamical friction. This proposal was motivated by the work
of
\citet{tou02}, which suggested that even a small fraction of mass in
retrograde orbits could excite a linear lopsided instability.

Counter--rotating streams of matter in a self--gravitating disc are
known to be unstable to lopsided modes \citep{zh78, ara87, saw88,
ms90, pp90, sm94, ljh97}. The dynamics of galactic
nuclei involve nearly Keplerian systems of stars or other
collisionless matter \citep{rt96, st99, sst99, tre01, ttk09}.
\citet{tou02} considered a {\it softened gravity} version of
Laplace--Lagrange theory of planetary motions, and showed that a
nearly Keplerian axisymmetric disc is linearly unstable to a $m = 1$
mode when even a small fraction of the disc mass is
counter--rotating. Softened gravity was introduced by \citet{mil71} to
simplify the analysis of the dynamics of stellar systems. In this form
of interaction, the Newtonian $1/d$ gravitational potential is
replaced by $1/\sqrt{d^2 + b^2}$, where $b>0$ is called the {\it
softening length}. In the context of waves in discs, it is well--known
that the softening length mimics the epicyclic radius of stars on
nearly circular orbits. Therefore, a disc composed of cold
collisionless matter interacting via softened gravity provides a
surrogate for a ``hot'' collisionless disc.

The goal of this paper is to formulate and analyse the
counter--rotating instability in the simplest model of a ``hot''
nearly Keplerian disc of collisionless self--gravitating matter.  To
this end we make the following choices: (i) The discs are assumed to
be made of matter whose self--interaction is through softened gravity;
(ii) A Wentzel--Kramers--Brillouin (WKB) analysis is made of the
linearised equations governing the perturbations. The unperturbed
two--component nearly Keplerian disc is introduced in \S~2 and the
apse precession rate is defined. The equations governing the
linearised perturbations and relevant potential theory for softened
gravity is given in \S~3.  The local (or WKB) approximation and
dispersion relation for local modes is derived in \S~4. This is used
to discuss stability, instability and overstability.
\S~5 considers the construction of global  modes for the case of a Kuzmin disc with 
equal masses in the counter--rotating components, and we conclude in \S~6.

\section{The Unperturbed Two--Component Disc}

We consider a disc of mass $M_d$ orbiting a central mass $M$. We
specialise to the nearly Keplerian case, $M_d\ll M\,$. Therefore the
force on the disc material is mostly Newtonian, giving rise to a near
equality between the frequencies of azimuthal and radial
oscillations. Test particle orbits may be thought of as osculating
Keplerian ellipses, whose apsides precess due to the self--gravity of
the disc, at rates that are smaller than the orbital frequency by a
factor $\sim\varepsilon \,=\, M_d/M\,\ll 1\,$.  Other forces could
also be responsible for the evolution of the disc over similar {\em
slow} time scales. We consider a cold collisionless disc, composed of
particles orbiting on two counter--rotating streams. The disc
particles interact with each other through softened gravity. The
central mass and the disc attract each other through (unsoftened)
Newtonian gravity. Our notation closely follows \citet{tre01}.

We employ polar coordinates ($r, \phi$) in the disc plane and place
the central mass at the origin. The unperturbed components are
axisymmetric with surface densities $\Sigma_d^+(r)\,\geq\,0\,$ and
$\Sigma_d^-(r)\,\geq\,0\,$, and circular velocities
$\bfv_d^{\pm}\,=\,\pm r\Omega(r)\hat{\phi}\,$, respectively. The
angular frequency, $\Omega(r)\,>\,0\,$, is determined by the total
gravitational potential:
\beq 
\Phi(r)\;=\;-\frac{GM}{r} \;+\;
\Phi_d(r) \;+\; \Phi_e(r)\,, \label{pot}
\eeq
\noindent where $\Phi_d$ is the self gravity of the disc
determined by the total surface density
$\Sigma_d(r)\,=\,\Sigma_d^+\,+\,\Sigma_d^-\,$, and $\Phi_e$ is a
non--Keplerian potential due to an external source. $\Phi_d$ is
$O(\varepsilon)$ compared to $GM/r\,$; we assume that $\Phi_e$ is also
$O(\varepsilon)\,$. Test particles on nearly circular prograde orbits
have azimuthal and radial frequencies, $\Omega\,>\,0$ and
$\kappa\,>\,0\,$; particles on nearly circular retrograde orbits have
frequencies, $-\Omega\,<\,0$ and $-\kappa\,<\,0\,$. The frequencies
are given by,
\begin{eqnarray} 
\Omega^2(r) &\;=\;&
\frac{GM}{r^3} \;+\;
\frac{1}{r}\frac{d}{dr}\,\left(\Phi_d\,+\,\Phi_e\right)\,,
\label{omega}\\[1em] 
\kappa^2(r) &\;=\;& \frac{GM}{r^3} \;+\;
\left(\frac{d^2}{dr^2}\,+\,\frac{3}{r}\frac{d}{dr}\right)
\left(\Phi_d\,+\,\Phi_e\right)\,.
\label{kappa}
\end{eqnarray}
\noindent The precession rate of the apsides of a nearly circular
orbit of angular frequency $\pm\Omega(r)\,$, is given by
$\pm\prate\,$, where,
\begin{eqnarray} 
\prate(r) &\;=\;& \Omega \;-\; \kappa\nonumber\\[1em]
&\;=\;&
-\frac{1}{2\Omega(r)}\left(\frac{d^2}{dr^2}
\,+\,\frac{2}{r}\frac{d}{dr}\right)
\left(\Phi_d\,+\,\Phi_e\right) \;+\; O(\varepsilon^2)\,.
\nonumber\\
&&\label{prate}
\end{eqnarray}
\noindent 

\section{Linear Response}

Let $\bfv_a^{\pm} \,=\, u_a^{\pm}(\bfr, t)\hat{r} \,+\,
v_a^{\pm}(\bfr, t)\hat{\phi}$ and $\Sigma_a^{\pm}(\bfr, t)$  be
infinitesimal perturbations to $\bfv_d^{\pm}$ and $\Sigma_d^{\pm}$. 
They satisfy the linearised Euler and continuity equations, 
appropriate to a cold disc:
\begin{eqnarray}
&&\frac{\p \bfv_a^{\pm}}{\p t} \;+\;
\left(\bfv_d^{\pm}\cdot\del\right)\bfv_a^{\pm} \;+\; 
\left(\bfv_a^{\pm}\cdot\del\right)\bfv_d^{\pm}  \;=\; -\del\Phi_a\,,
\label{euler}\\[1em]
&&\frac{\p \Sigma_a^{\pm}}{\p t} \;+\; \del\cdot\left(\Sigma_d^{\pm}
\bfv_a^{\pm} \,+\, \Sigma_a^{\pm}\bfv_d^{\pm}\right) \;=\; 0\,,
\label{cont}
\end{eqnarray}
\noindent where $\Phi_a(\bfr, t)$ is the perturbing potential. We
write the variables $(\Sigma_a^{\pm}, u_a^{\pm}, v_a^{\pm}, \Phi_a)$
in the form $X_a(r, \phi, t) \,=\, X_a^m(r)\exp{[i(m\phi -
\omega t)]}\,$. Substituting these in equations~(\ref{euler}) and
(\ref{cont}), straightforward manipulations give,    
\begin{eqnarray}
&&u_a^{m\pm} \;=\; -\frac{i}{D_m^{\pm}}\left[ (\pm m\Omega -
\omega)\frac{d}{dr} \,\pm\,
\frac{2m\Omega}{r}\right]\Phi_a^m\,,
\label{pertu}\\[1em]
&&v_a^{m\pm} \;=\;
\frac{1}{D_m^{\pm}}\left[\pm\,\frac{\kappa^2}{2\Omega}\,\frac{d}{dr}
\,+\, \frac{m}{r}(\pm m\Omega - \omega)\right]\Phi_a^m\,,
\label{pertv}\\[1em]
&&i(\pm m\Omega - \omega)\Sigma_a^{m\pm} \;=\;
-\frac{1}{r}\frac{d}{dr} \left(r\Sigma_d^{\pm}u_a^{m\pm}\right) \;-\;
\frac{im}{r}\Sigma_d^{\pm} v_a^{m\pm}\,,
\nonumber\\
&&\label{pertcont}
\end{eqnarray}
\noindent where 
\beq
D_m^{\pm} \;=\; \kappa^2 \;-\; (\pm m\Omega - \omega)^2\,.
\label{dm}
\eeq
\noindent Equations~(\ref{pertu})---(\ref{dm}) give the 
linear responses of the surface densities and velocities of the two
components, to a specified perturbing potential, $\Phi_a\,$.

For a self--consistent response, the perturbing potential,
$\Phi_a^m(r)$, depends only on the total surface density,
$\left(\Sigma_a^{m+}\,+\,\Sigma_a^{m-}\right)$; in fact this is the
only coupling between the two counter--rotating components. The
Poisson integral is,
\beq
\Phi_a^m(r) \;=\; \int_0^{\infty} dr'\,r'\,P_m(r,
r')\,\left[\Sigma_a^{m+}(r') \,+\,\Sigma_a^{m-}(r')\right]\,. 
\label{poisson}
\eeq
\noindent The unperturbed disc potential, $\Phi_d$, is related to
the unperturbed surface density, $\Sigma_d \,=\,
\left(\Sigma_d\,+\,\Sigma_d\right)$ through equation~(\ref{poisson}), 
when $m\,=\,0$. The kernel,
\beq
P_m(r, r') \;=\; -\frac{\pi G}{r_>}B_1^m(\alpha, \beta) \;+\;
\pi G\frac{r}{r'^2}\left(\delta_{m1} \,+\, \delta_{m, -1}\right)\,,
\label{kernel} 
\eeq
\noindent includes direct and indirect contributions. Here 
$r_<\,=\, {\rm min}(r, r')$, $r_>\,=\, {\rm max}(r,
r')$, $\alpha \,=\, r_</r_>\,$, $\beta \,=\, b/r_>\,$,
and, 
\beq
B_s^m(\alpha, \beta) \;=\; \frac{2}{\pi}\int_0^{\infty}
\frac{d\theta\,\cos{m\theta}}{\left(1 \,-\, 2\alpha\cos{\theta}
\,+\, \alpha^2 \,+\, \beta^2\right)^{s/2}}\;,
\label{laplace}
\eeq
\noindent is a generalisation of the Laplace coefficients,
introduced by \citet{tou02}; in the limit of no softening,
$B_s^m(\alpha, 0) \,=\, b_{s/2}^m(\alpha)$, the Laplace coefficients
familiar from celestial mechanics
\citep{md99}. Equations~(\ref{pertu})---(\ref{laplace}) determine the
self--consistent, linear modes of axisymmetric discs, whose
counter--rotating components interact through softened gravity.

\section{The Local Approximation}
\label{sec:localapprox}

A perturbation, $X_a^m(r) \,=\, \left|X_a^m(r)\right|\exp{[i\int^r
dr\,k(r)]}\,$, is referred to as tightly--wound, if the radial
wavenumber is large in the sense, $|rk(r)|\gg |m|$. To leading order
in $|m/rk(r)|$, the WKB approximations to the linear responses of
equations~(\ref{pertu})---(\ref{pertcont})  are, 
\begin{eqnarray}
u_a^{m\pm}(r) &\;=\;& \frac{(\pm m\Omega
-\omega)}{D_m^{\pm}}\:k\Phi_a^m \,,\label{wkbu}\\[1em] v_a^{m\pm}(r)
&\;=\;& \pm i\,\frac{\kappa^2}{2\Omega D_m^{\pm}}\:k\Phi_a^m
\,,\label{wkbv}\\[1em]
\Sigma_a^{m\pm}(r) &\;=\;&
-\,\frac{\Sigma_d^{\pm}}{D_m^{\pm}}\:k^2\Phi_a^m \,.\label{wkbcont}
\end{eqnarray}
\noindent The responses are singular at radii, where $D_m^{\pm}
\,=\,0$. For a self--consistent $\Phi_a^m$, the potential theory
of the previous Section simplifies in the WKB limit \citep{mil71}, 
\beq 
\Phi_a^m \;=\; -\,2\pi G\:\frac{\exp{\left(-|k|b\,\right)}}{|k|}\,
\left(\Sigma_a^{m+} \,+\,\Sigma_a^{m-}\right)\,.
\label{softpot} 
\eeq 
\noindent Substituting for $\Sigma_a^{m+}$ and $\Sigma_a^{m-}$ from 
equation~(\ref{wkbcont}), and eliminating $\Phi_a^m$, gives the WKB
dispersion relation,
\beq
D_m^+\,D_m^- \;=\; 2\pi G\,|k|\exp{\left(-|k|b\,\right)}\left[
D_m^-\Sigma_d^+ \,+\, D_m^+\Sigma_d^-\right]\,,
\label{disprel}
\eeq
\noindent which is a quartic equation in $\omega$. All the 
coefficients of the various powers of $\omega$ being real, if $\omega$
is a solution, so is its complex conjugate, $\omega^*$. It should be
noted that we have not made any assumptions about the Keplerian nature
of the disc. Therefore the dispersion relation of
equation~(\ref{disprel}) is valid for a non Keplerian disc as well,
with or without the central mass. In general, $\omega$ will not be
small, compared to either $\Omega$ or $\kappa$, so the the dispersion
relation is not restricted to {\sl slow} perturbations. When the
counter--rotating component is absent (i.e. $\Sigma_d^- \,=\, 0$),
equation~(\ref{disprel}) reduces to the WKB dispersion relation,
familiar from Problem~(6--5) of
\citet{bt87}.     

We recall the result for the stability of axisymmetric ($m=0$) perturbations.
When $m\,=\,0$, we have $D_0^+ \,=\, D_0^+ \,=\, \kappa^2 \,-\, 
\omega^2$, and equation~(\ref{disprel}) becomes, 
\beq
\omega^2 \;=\; \kappa^2 \;-\; 2\pi G\Sigma_d 
|k|\exp{\left(-|k|b\,\right)}\,,
\label{dr0}
\eeq
\noindent which is identical to the dispersion relation for 
a disc without counter--rotating components. It is straightforward to
prove (as Problem~(6--5) of \citet{bt87} invites the reader to) that
the disc is stable to short--wavelength axisymmetric perturbations, if
\beq
b \;>\; b_0 \;\equiv\; \frac{2\pi G\Sigma_d}{\kappa^2 {\rm e}}\,.
\label{stab0}
\eeq 

\subsection{Slow  \mbox{\boldmath $m\,=\,1$} Perturbations}

When the azimuthal wavenumber is $m\,=\,1$, the near equality between
$\Omega$ and $\kappa$---see equations~(\ref{omega}) and
(\ref{kappa})---enables slow modes, for which $|\omega/\Omega| \,=\,
O(\varepsilon)\ll 1$. It is convenient to define,
\beq
\Sigma_d^+ \;=\; (1 - \eta)\Sigma_d\,,\qquad \mbox{and}\qquad
\Sigma_d^- \;=\;\eta\Sigma_d\,,
\eeq
\noindent where $0\leq\eta(r)\leq 1$ is the local mass fraction in
the  counter--rotating component. We also introduce a local frequency,
\beq
\cals(r, |k|)  \;=\; \frac{\pi G\Sigma_d(r)}{\Omega(r)}\,|k|
\exp{\left(-|k|b\,\right)}\,, 
\label{sdef}
\eeq 
\noindent whose maximum value at any $r$, 
\beq
\cals_{\rm max}(r) \;=\; \frac{\pi G \Sigma_d}{{\rm e} b \Omega}
\;=\; \frac{\kappa^2\,b_0}{2\Omega b} \;\simeq\;          
\frac{\Omega b_0}{2b}\,,
\label{smax}
\eeq 
\noindent is attained for $|k| = 1/b\,$. Here $b_0$ is the minimum 
softening length, defined in equation~(\ref{stab0}), that ensures
local stability to axisymmetric perturbations.

From equations~(\ref{prate}) and (\ref{dm}), we have,
\begin{eqnarray}
D_1^+ &\;=\;& 2\Omega\left(\omega \,-\, \prate\right)
\;+\; O(\varepsilon^2)\,,\label{dm1+}\\[1em]
D_1^- &\;=\;& -\,2\Omega\left(\omega \,+\, \prate\right)
\;+\; O(\varepsilon^2)\,.\label{dm1-}
\end{eqnarray}
\noindent When these are substituted in equation~(\ref{disprel}), 
a little rearrangement provides the dispersion relation for slow, $m
\,=\, 1$ perturbations:
\beq
\omega^2 \;-\; \cals\left(1 \,-\, 2\eta\right)\omega \;-\;
\prate\left(\prate \,+\, \cals\right) \;=\; 0\,,
\label{dr1}
\eeq
\noindent whose solution is,
\beq 
\omega \;=\; \frac{\cals}{2}\left(1 \,-\, 2\eta\right)
\;\pm\; \frac{1}{2}\sqrt{\cals^2\left(1 \,-\, 2\eta\right)^2
\,+\, 4\prate\left(\prate \,+\, \cals\right)}\,.
\label{dr1soln}
\eeq 
\noindent Equations~(\ref{dr1}) and (\ref{dr1soln}) are invariant
under $(\eta, \omega)\to(1-\eta, -\omega)$, because this operation is
equivalent to interchanging the meaning of ``prograde'' and
``retrograde''. It is convenient to first consider two special cases:

\noindent (i) {\sl No counter--rotation, $\eta \,=\,0$}:
When $\eta \,=\, 0$, equation~(\ref{dr1soln})
admits the two roots, $\omega \,=\, \prate + \cals$, and $\omega
\,=\, -\prate$. The former root corresponds to the eqn.~(14)
of \citet{tre01}, and implies that the disc is locally stable to all
$m \,=\, 1$ perturbations. However, $\omega \,=\, -\prate$ is a
spurious solution, arising from multiplication by $D_1^-$ in the
derivation of equation~(\ref{disprel}). Henceforth we assume that
$\eta\neq 0$.

\noindent (ii) {\sl Equal counter--rotation, $\eta \,=\,1/2$}:
When there is equal mass (locally) in the prograde and retrograde 
components, the two roots of equation~(\ref{dr1soln}) are, 
$\omega \,=\, \pm\sqrt{\prate\,(\prate\,+\,\cals)}$. If $\prate$
happens to be positive, then $\omega$ is real, and the disc is 
locally stable. However, $\prate < 0$ for most continuous discs, 
hence $\omega$ can be either real, or purely imaginary;  
there is no local {\sl overstability}. The criteria for (in)stability
are discussed below, along with the case of general $\eta$.

The sign of the discriminant of equation~(\ref{dr1soln}),
\beq
\cald \;=\; \cals^2\left(1 \,-\, 2\eta\right)^2
\,+\, 4\prate\left(\prate \,+\, \cals\right)\,,
\label{discr}
\eeq
\noindent determines whether $\omega$ is real, or complex. 
If $\prate > 0$, then $\cald > 0$, hence $\omega$ is real. 
However, $\prate < 0$ for most continuous discs, and it 
is straightforward to determine that $\cald < 0$,
if $\cals$ lies in the range of values, $0 < \cals_{-} < \cals <
\cals_{+}$,  where 
\beq
\cals_{\pm} \;=\; \frac{2|\prate|}{(1-2\eta)^2} 
\left[1 \,\pm\, 2\sqrt{\eta\,(1 -\eta)}\right]\,.
\label{srange}
\eeq
\noindent However, we noted earlier that the maximum value 
that $\cals$ can take is $\cals_{\rm max}$, given by
equation~(\ref{smax}). Hence, at a specified $r$, $\cald$ is positive 
for all $k$, if $\cals_{-} > \cals_{\rm max}$. Therefore, the disc is
stable to all shortwavelength $m \,=\, 1$ perturbations, if 
\begin{eqnarray}
\nonumber b \;&>&\; b_1 \;\equiv\; \frac{\pi G\Sigma_d}{e\Omega|\prate|}\,
\left[\frac{(1-2\eta)^2}{2 \,-\, 4\sqrt{\eta\,(1 -\eta)}}\right]\\
\;&\simeq&\; b_0\,\left(\frac{\Omega}{2|\prate|}\right)\,
\left[\frac{(1-2\eta)^2}{2 \,-\, 4\sqrt{\eta\,(1 -\eta)}}\right]\,.
\label{b1}
\end{eqnarray}
\noindent The $\eta$--dependent factor in equation~(\ref{b1}) does
not vanish for any value of $\eta$. In fact, as $\eta\to 1/2$, 
the term in $[\;]$ approaches unity, giving the stability criterion
for the case $\eta = 1/2$, as may be verified independently.
Since $(|\prate|/\Omega) = O(\varepsilon)\ll 1$, the $b$ required for
local stability, according to equation~(\ref{b1}), equals
$b_0$ (which is the minimum  softening required for local
axisymmetric stability) multiplied by a large factor, of order
$1/\varepsilon$. To the extent softening mimics ``heat'' (more
precisely, the epicyclic radius) in collisionless discs, this 
criterion suggests that a Keplerian disc would have to
very hot indeed, to be able to avoid a local instability to $m
\,=\, 1$ perturbations. Hence one is led to consider overstable
perturbations. 

Overstability occurs when $\cald < 0$. We write $\omega \,=\,
\Omega_p \,\pm\, i\Gamma$, where  $\Omega_p$ is the pattern speed
of the $m\,=\, 1$ perturbation, and $\Gamma > 0$ is the growth rate.
From equation~(\ref{dr1soln}),  
\begin{eqnarray}
\Omega_p &\;=\;& \frac{\cals}{2}\left(1 \,-\, 2\eta\right)\,,
\label{patsp}\\[1em]
\Gamma &\;=\;& \frac{1}{2}\sqrt{\left|\cals^2\left(1 \,-\,
2\eta\right)^2 \,+\, 4\prate\left(\prate \,+\, \cals\right)\right|}\,.
\label{growth}
\end{eqnarray}

\section{Global unstable modes for equal counter--rotation}

We have seen in the previous section that in the case of equal counter--rotation
(i.e. $\eta=1/2$) local analysis predicts purely unstable {\it slow $m=1$ modes}. We will
now go beyond the local analysis and construct global $m=1$ WKB modes for this case. 
To do this it is necessary to consider a concrete example. It is useful to
take the axisymmetric unperturbed disc to be a Kuzmin disc because 
(i) the Kuzmin disc is centrally concentrated and is hence a plausible candidate
for being a quite generic case; (ii) the surface density, the self--gravitational potential and 
the precession rate all have explicit analytical forms; (iii) the slow modes of the Kuzmin disc were
studied in \citet{tre01} for the case of no counter--rotation.

The surface density of the Kuzmin disc is given by,
\begin{equation}
\Sigma_d(r) = \frac{aM_d}{2\pi(r^2+a^2)^{5/2}}\,,
\end{equation}
where $M_d$ is the disc mass and $a$ is the central concentration
parameter. The precession rate due to the Kuzmin disc is given by
\begin{equation}
{\prate}_d = -\frac{3GM_da^2}{2\Omega(r)(r^2+a^2)^{5/2}}\,,
\end{equation}
where the rotational frequency is given by the Keplerian flow due to
the central mass $M$ as $\Omega(r) = \sqrt{GM/r^3}$. Let us first
consider the stability of the disc under axisymmetric
perturbations. In section \S~\ref{sec:localapprox} we derived the
minimum value of the softening parameter $b_0$ that ensures local
stability. In the slow mode limit $\kappa(r)=\Omega(r)$ to the zeroth
approximation. Therefore we find
\begin{equation}
b_0 = \frac{2\pi G\Sigma_d}{\kappa^2{\rm e}} = \frac{aM_d}{{\rm e}M} \frac{r^3}{(r^2+a^2)^{3/2}} \,.
\end{equation}
The largest value of $b_0 = aM_d/{\rm e}M$, therefore the smallest softening parameter $b$
that ensures stability everywhere satisfies $b/a > M_d/{\rm e}M_0$. Let us define a parameter
$R=b/a$ and a parameter $\beta$ through  $M_d/M_0 = \beta R $.  In terms of these parameters the previous inequality
gives $\beta < \rm e$. Substituting $\eta = 1/2$ in equation~(\ref{dr1soln}), we get
\beq
\cals = \frac{\omega^2-\prate^2}{\prate}\,.
\eeq 
Now we substitute for $\prate$ and recall that we are looking for global
unstable modes. Therefore $\omega^2=-\Gamma^2$, and
\begin{equation}
a|k| \exp{\left(-|k|b\right)} = \frac{4}{3}
\left(\frac{M_0}{M_d}\right)^2 \frac{\Gamma^2}{\Omega(a)^2}
\frac{(\xs^2+1)^4}{\xs^3}+\frac{3}{1+\xs^2}\,,
\end{equation}
where $\xs=r/a$. By defining $\ks = a|k|$ and $\gs = \sqrt{4/3}\Gamma/\beta\Omega(a)$ we finally obtain
\begin{eqnarray}
R|\ks| \exp{\left(-|\ks|R\right)} &=& \Phi(\xs;\gs,R)\,,\quad\mbox{where}\nonumber\\[1em]
\Phi(\xs;\gs,R) &=&
\frac{\gs^2}{R} \frac{(\xs^2+1)^4}{\xs^3}+\frac{3R}{1+\xs^2}
\label{eq:wkb_numeric}
\end{eqnarray}

\subsection{Numerical Results}

Global unstable modes are determined by numerically solving
equation~(\ref{eq:wkb_numeric}) for $\ks$ for a given value of $\gs$
and $R$ and applying a quantization condition to obtain growth
rate. Note that the right hand side of equation~(\ref{eq:wkb_numeric})
blows up at $\xs=0$ and $\xs
\rightarrow
\infty$; however it is bounded from below and has a minimum at $\xs
\simeq 1$, the exact value depending on the precise value of $\gs$ and $R$. 
Since the left hand side of this equation has a maximum value equal to
unity at $\ks=1$, the equation does not admit any solutions if
$\min\left[\Phi(\xs;\gs,R)\right] > 1$. However, for $\min\left[\Phi(\xs;\gs,R)\right] <
1$, the equation admits two real roots, one each on either side of
$\ks=1$. We denote the roots with $\ks<1$ as the long-wavelength
branch and the one with $\ks>1$ as the short-wavelength branch.

As noted above, the right hand side of equation~(\ref{eq:wkb_numeric})
is unbounded from above and blows up at small and large values of
$\xs$. It is clear that the real roots exist only for a finite range
of the radial coordinate, $a<\xs<b$, where both $a$ and $b$ depend on
the parameters $\gs$ and $R$ in a complicated manner. We shall assume
that at these points the wave is reflected, and we therefore impose
the Bohr--Sommerfeld quanitization condition
\begin{equation}
\int_a^b \ks d\xs = \left(n+\half\right)\pi\,.
\end{equation}

\begin{figure}
\centering
\includegraphics[width=0.5\textwidth]{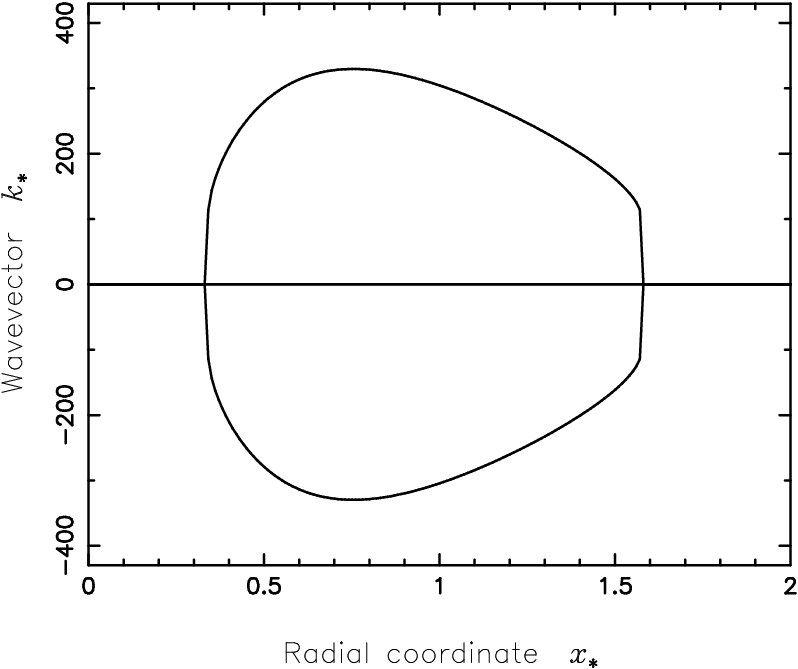}
\caption{Phase plot of a possible mode in the short--wavelength branch for $R=0.01$ and $\gs=0.0001$.
}
\label{fig1}
\end{figure}

\begin{figure}
\centering
\includegraphics[width=0.5\textwidth]{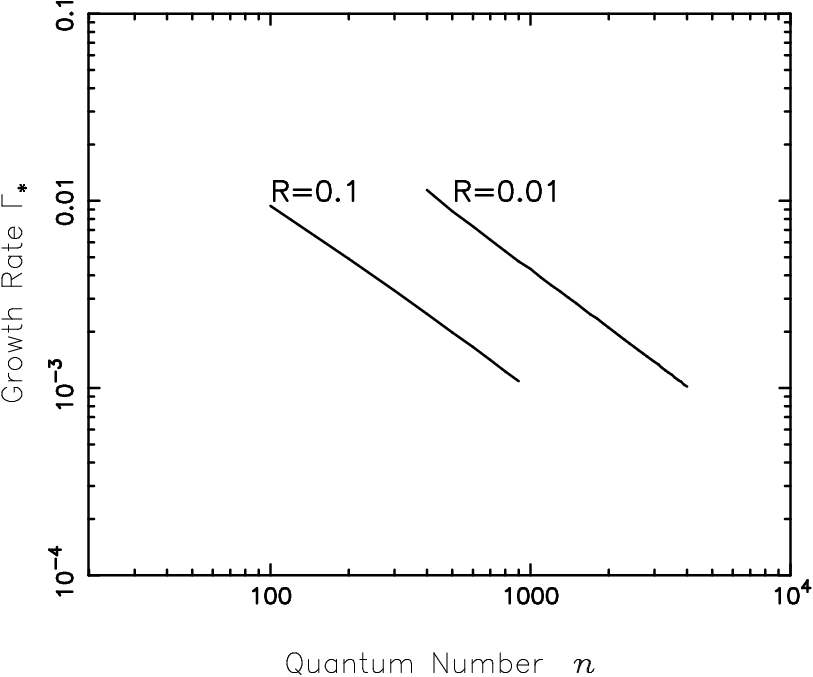}
\caption{Dimensionless growth rate versus quantum number of the mode for the 
short--wavelength branch.
}
\label{fig2}
\end{figure}

\begin{figure}
\centering
\includegraphics[width=0.5\textwidth]{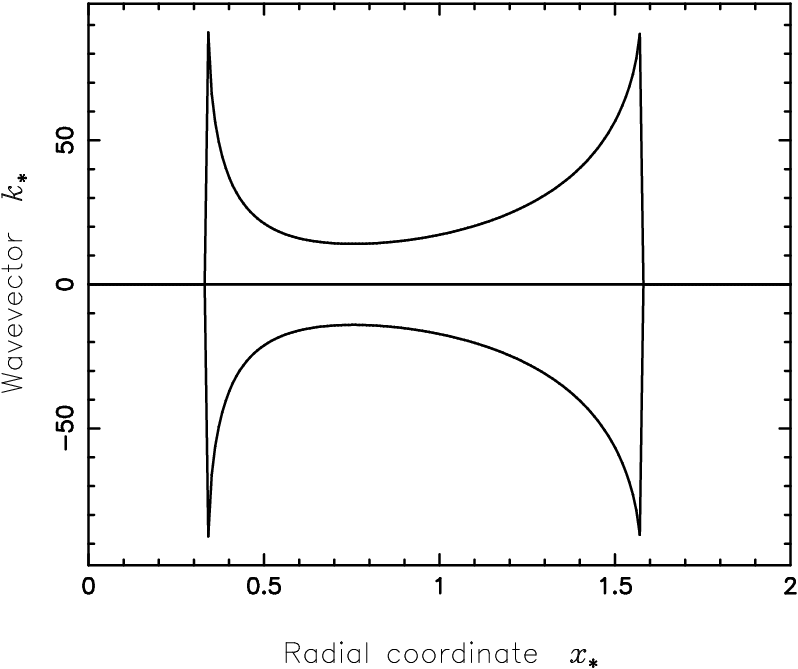}
\caption{Phase plot of a possible mode in the long--wavelength branch for $R=0.01$ and $\gs=0.0001$.
}
\label{fig3}
\end{figure}

\begin{figure}
\centering
\includegraphics[width=0.5\textwidth]{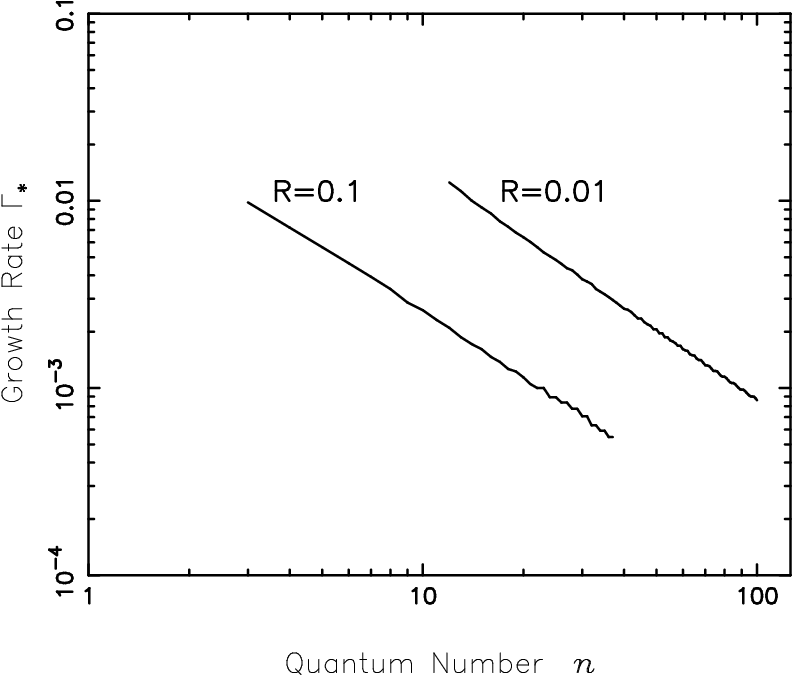}
\caption{Dimensionless growth rate versus quantum number of the mode for the 
long--wavelength branch.
}
\label{fig4}
\end{figure}

We find that the quantization condition cannot be satisfied for $R
> 1$. However, $R=b/a > M_d/{\rm e}M \ll 1$; therefore a small value
of $R$ is allowed by the inequality. We seek the global modes for two
cases; $R=0.1$ and $R=0.01$. In Figure~(\ref{fig1}) and
~(\ref{fig3}) we display the phase plots for the short and long wavelength
branches for $R=0.01$ and $\gs=0.0001$. The general trend is that as
$\gs$ increases the contours become narrower in the horizontal
direction and expand in the vertical direction in such a manner that
the area under the closed curve decreases as $\gs$ increases. This
can be seen clearly in Figures~(\ref{fig2}) and (\ref{fig4}), where we
have plotted the spectrum for the short and long wavelength branches,
where we see that larger quantum numbers correspond to smaller values
of the growth rate. The two figures show that the effect of changing
$R$ is to translate the spectrum horizontally. This is only an
approximate behaviour and does not imply anything special. More
signicantly the spectrum is seen to be very close to a power law. This
behaviour is quite robust and persists for other values of $R$.

\section{Conclusions}

The principal aim of this work is to present the simplest model that permits a largely analytical
exploration of the $m=1$ counter--rotating instability in a ``hot'' nearly Keplerian disc of collisionless 
self--gravitating matter. To this end we have considered a two--component softened gravity disc, 
and performed a linearised WKB analysis of both local and global modes. We derive an analytical expression 
for local WKB waves for arbitrary $m$, which turns out to be quartic in the frequency $\omega$. 
Specialising to $m=1$, we show that $\omega$ is smaller than the (Keplerian) orbital frequency by
the small quantitity $\varepsilon = M_d/M$ (the ratio of the disc mass to the mass of the central object);
in other words, the $m=1$ modes are {\it slow modes}. The dispersion relation now reduces to a quadratic 
equation in $\omega$. Hence the criteria for stability, instability and overstability can be readily derived in 
simple analytical forms. For a one--component disc (which does not have any counter--rotation), the $m=1$
modes are stable, consistent with the results of \citet{tre01}. Equal mass in the two counter--rotating 
components corresponds to the case of not net rotation. In this case we find that 
the local modes are purely unstable (i.e. not overstable), consistent with \citet{ara87, pp90, sm94, ljh97, tou02}.
However the general case of arbitrary mass ratio in the two counter--rotating components corresponds to
overstability, and we show analytically that the discs must be unrealistically hot to avoid an overstability.
We finally contructed global  WKB modes, numerically,  for the case of a Kuzmin disc for the case of no net rotation, by 
using Bohr--Sommerfeld quantisation.

\def\etal{{\it et~al.}}
\def\apj{{Astroph.\@ J. }}
\def\mnras{{Mon.\@ Not.\@ Roy.\@ Ast.\@ Soc.}}
\def\aap{{Astron.\@ Astrophys.}}
\def\aj{{Astron.\@ J.}}
\def\apjl{{Astrophysical.\@ J. {\rm Letters}}}
\def\apss{{Astroph.\@ Space \@ Science}}
 
\end{document}